\documentclass[journal]{IEEEtran}
\ifCLASSINFOpdf
\else
\fi

\usepackage{graphicx}
\graphicspath{{figures/}}
\DeclareGraphicsExtensions{.pdf,.jpg,.png}

\usepackage{adjustbox}	% - Compressing tables
\usepackage{booktabs}
\usepackage{multirow}
\usepackage{makecell}

\usepackage{amsfonts}
\usepackage{caption}
\usepackage{subcaption}

\usepackage{ntheorem}
\theoremseparator{:}
\newtheorem{hyp}{Research Question}

\usepackage{xcolor,colortbl}
\newcommand{\review}[1]{#1}

\begin{document}
\title{Identifying Student Profiles within Online Judge systems using Explainable Artificial Intelligence}

\author{Juan Ram{\'o}n Rico-Juan,
        V{\'i}ctor M. S{\'a}nchez-Cartagena,
        Jose J. Valero-Mas,
        and Antonio Javier Gallego% <-this % stops a space
\thanks{The authors are with the Department of Software and Computing Systems, University of Alicante, 03690, Alicante Spain (e-mail:juanramonrico@ua.es; \{vmsanchez, jjvalero, jgallego\}@dlsi.ua.es).}
\thanks{Corresponding author: Antonio Javier Gallego (e-mail: jgallego@dlsi.ua.es)}% <-this % stops a space

}


\markboth{IEEE TRANSACTIONS ON LEARNING TECHNOLOGIES, VOL. XX, NO. X, XXXXXX 20XX}%
{Shell \MakeLowercase{\textit{et al.}}: Bare Demo of IEEEtran.cls for IEEE Journals}



\maketitle

\begin{abstract}
Online Judge (OJ) systems are typically considered within programming-related courses as they yield fast and objective assessments of the code developed by the students. Such an evaluation generally provides a single decision based on a rubric, most commonly whether the submission successfully accomplished the assignment. Nevertheless, since in an educational context such information may be deemed insufficient, it would be beneficial for both the student and the instructor to receive additional feedback about the overall development of the task. This work aims to tackle this limitation by considering the further exploitation of the information gathered by the OJ and automatically inferring feedback for both the student and the instructor. More precisely, we consider the use of learning-based schemes---particularly, Multi-Instance Learning and classical Machine Learning formulations---to model student behaviour. Besides, Explainable Artificial Intelligence is contemplated to provide human-understandable feedback. The proposal has been evaluated considering a case of study comprising 2,500 submissions from roughly 90 different students from a programming-related course in a Computer Science degree. The results obtained validate the proposal: the model is capable of significantly predicting the user outcome (either passing or failing the assignment) solely based on the behavioural pattern inferred by the submissions provided to the OJ. Moreover, the proposal is able to identify prone-to-fail student groups and profiles as well as other relevant information, which eventually serves as feedback to both the student and the instructor.
\end{abstract}


% \fi
\IEEEpeerreviewmaketitle

\section{Introduction}
\label{sec:introduction}


% Introducing the online judges and their application in programming courses
\IEEEPARstart{O}{riginally} coined by \cite{kurnia2001online}, the term Online Judge (OJ) denotes those systems devised for the automated evaluation and grading of programming assignments, \review{which usually take the form of} 
%More precisely, these methods stand for the 
online evaluation services capable of collecting source codes, compiling them, assessing their results, and computing scores based on different criteria~\cite{wasik2018survey}. These automated tools have been particularly considered in two precise, yet related, scenarios~\cite{yera2017recommendation}: (i) programming contests and competitions, and (ii) educational contexts in academic degrees. This work focuses on the latter scenario, in particular, on programming courses from Computer Science studies in higher education institutions.

% Effort , error-prone...
OJ systems are successful in the education field because they overcome the main issues associated with the manual evaluation of assignments~\cite{CheangKurniaLimOon:CE:2003}: in opposition to human grading, which is deemed as a tedious and error-prone task, these tools provide immediate corrections of the submissions regardless of the number of participants. Moreover, the competitive learning framework that these schemes entail proves to benefit \review{the success of the learning process~\cite{ReguerasVerduMunozdeCastroVerdu:IEEEToE:2009}.}
% ,pingwang2016ojpot
%besides improving the logical thinking, research, and innovation capabilities of the students~\cite{wang2017design}.

% Introducing their inherent limitations for providing feedback to the student
Despite their clear advantages, OJ systems do not provide the student nor the instructor with any feedback from the actual submission apart from whether the provided code successfully accomplished the assignment~\cite{Mani:ICCSE:2014}. However, the information gathered by the OJ system may be further exploited to enrich the educational process by automatically extracting additional insights such as student habits or patterns of behaviour related to the success (or failure) of the task. In this regard, one may resort to the so-called Educational Data Mining (EDM), 
%paradigm that considers the application of data mining methods to exploit and gather knowledge from educational data~\cite{AsifMerceronAliHaider:CE:2017}.
%Nevertheless, the main drawback in educational OJ systems is that the assessment is commonly reduced to a correct/incorrect grading~\cite{Mani:ICCSE:2014}. This particularity hinders the educational process since neither the student nor the instructor are provided with any feedback from the actual submission besides whether the provided code successfully accomplished the assignment. 
%However, it may be argued that the information gathered by the OJ system may be further exploited to automatically extract additional insights such as student habits or patterns of behaviour related to the success (or failure) of the task. In this regard, one may resort to the so-called Educational Data Mining (EDM) paradigm that considers the application of data mining methods to exploit and gather knowledge from educational data~\cite{AsifMerceronAliHaider:CE:2017}.
% Educational Data Mining as a way of finding descriptive patterns to characterise students; ML as a particular tool in EDM and the particular case of MIL for the current work
%Formally, EDM represents 
\review{a discipline meant to infer descriptive patterns and predictions from educational settings\review{~\cite{AsifMerceronAliHaider:CE:2017}}.}
%such as characterising behaviours and achievements of the learners, analysing domain knowledge content, autonomous assessments or educational functionalities, among others~\cite{PenaAyala:ESWA:2014}. 
Within this discipline, Machine Learning (ML) is reported as one of the main 
\review{enabling technologies}
%frameworks for carrying out this knowledge inference process 
due to its power and flexibility. Some success cases can be found in the work by~\cite{zhang2021examining}, devoted to assessing the performance of the instructor; the approach by~\cite{gray2019utilizing}, aimed at predicting student grades at an early stage; or the work by~\cite{rico2019automatic}, focused on detecting inconsistencies in peer-review assignments. 
%In our aforementioned context of educational OJ systems for programming courses, we consider the possibility of exploiting EDM tools to automatically provide feedback, both to the student and the instructor, about the assignments being developed.
\review{In this work, we apply EDM to automatically provide feedback about the assignments, both to the student and the instructor, in the context of OJ systems for programming courses.}

% Some of the information and meta-information that we can extract from this type of system include the number of submissions and their periodicity, whether they were successful or not, whether they were made close to the deadline, or how many days the students have worked.}

% Our case
% Providing feedback in OJ --- Identifying student profiles in OJ ---
When an OJ is used for grading a programming assignment, there is usually a time slot in which students can perform as many submissions as they want. The final grade of a student in the activity is typically computed from the best submission. During that time slot, data usually exploited in EDM, such as grades obtained in previous activities or course attendance~\cite{gray2019utilizing}, may not be available. Moreover, other data used to predict student performance, such as socioeconomic background or academic success in other courses\review{~\cite{alturki2021using}}, may not be usable from an ethical point of view due to the potential biases it would introduce.

\review{In spite of the lack of available data, it would still be desirable to be able to detect at-risk students before the assignment deadline. Thus, aided by the use of meta-information gathered from the submission process---\textit{e.g.}, the number of code submission attempts or the date of the first submission---we devised an EDM approach with two types of outcomes: (i) the success probability of a new student, and (ii) the identification of different student profiles to provide feedback to both the instructor and the student thyself. Note that such pieces of information may be used not only to prevent inadequate student attitudes by providing the appropriate observations about the development of the task but also to properly adjust the difficulty of the different assignments, among other possible corrective actions towards the success of the course.}

\review{Since the set of code submissions made by a student somehow characterises the student profile to be estimated, the problem may be modelled as a Multi-Instance Learning (MIL) task~\cite{foulds2010review}. This learning framework introduces the concept of \textit{bag}, \textit{i.e.}, a set with an indeterminate number of instances that is assigned a single label~\cite{zhang2009generalized}. MIL has been successfully considered in the EDM literature~\cite{AnupamaKumarVijayalakshmi:Book:2018}, as in the work by \cite{ZafraRomeroVentura:HEDM:2010}, which compares MIL against ML for predicting the student performance. In our case, each of these bags gathers the different code submissions made by each student, being labelled as either positive or negative depending on whether the student eventually passed the assessment by the OJ system.}

%However, it would still be beneficial to be able to detect different student profiles.
%It should be noted that in OJ evaluation systems it is not possible to precisely determine the student profile before the scheduled time for the assignment finishes. This is due to the fact that, during that period, the student may produce an undetermined number of submissions that may remarkably differ in the actual correctness of the task. 
%with the goal of inducing one or more concepts out of these collections and the elements within~\cite{zhang2009generalized}. MIL has been successfully considered in the EDM literature as in the work by \cite{ZafraRomeroVentura:HEDM:2010}, which compares MIL against ML for predicting the student performance, or that by \cite{ZafraVentura:ASOC:2012}, which extends the former one to the case of virtual learning environments. A thorough revision in terms of taxonomy and efficiency analysis can be found in the work by~\cite{AnupamaKumarVijayalakshmi:Book:2018}.

% Limitation of MIL -> not explainable; XAI
Nevertheless, the fact that both ML and MIL strategies generally work in a \textit{black box} manner hinders their application in this feedback-oriented context~\cite{KomarekBrabecSomol:ECML:2021}. In this regard, the field of Explainable Artificial Intelligence (XAI) is gradually gaining attention to tackle such limitation by devising methodologies that allow humans to understand and interpret the decisions taken by a computational model~\cite{BurkartHuber:JAIR:2021}. However, while XAI has been largely studied in the ML field, this has not been the case in the MIL one~\cite{arrieta2020explainable}. %In this regard, one may resort to existing mechanisms for mapping MIL problems into an ML framework to enable the use of XAI in these cases.

% Proposal
Considering all the above, this work presents a method to identify student profiles in educational OJ systems with the aim of providing feedback to both the students and the instructors about the development of the task. More precisely, the proposal exclusively relies on the meta-information extracted from these OJ systems and considers a MIL framework to automatically infer these profiles together with XAI methods to provide interpretability about the estimated behaviours. \review{In order to apply XAI to MIL problem, a novel policy for mapping the MIL representation to an ML one is proposed for the particular task at hand. } 
%and compared against the rest of schemes. 
The proposed methodology has been evaluated in a case of study comprising three academic years of a programming-related course with more than 2,500 submissions of two different assignments. For this, more than 20 learning-based strategies comprising ML, MIL, and MIL-to-ML mapping methods have been assessed and compared to prove the validity of the proposal. \review{The results obtained show that the proposal adequately models the user profile of the students while it also provides a remarkably precise estimator of their chances to succeed or fail in the posed task solely based on the meta-information of the OJ.}

% Roadmap
The rest of the work is organised as follows: Section~\ref{sec:context} reviews the related literature to contextualise the work; Section~\ref{sec:method} presents the proposed methodology; Section~\ref{sec:study_case} introduces the case of study examined; Section~\ref{sec:setup} details the experimental set-up considered; Section~\ref{sec:results} shows and discusses the results obtained; Section~\ref{sec:discussion} summarises the insights obtained in the work; and finally, Section~\ref{sec:conclusions} concludes the work and outlines future research line to address.

%---------------------------------------------------------------------------------------------------------
\section{Related Work}
\label{sec:context}

This section reviews the literature related to the proposed approach for assessing the performance of the students when considering OJ systems in programming courses. In this regard, we first revise the historical usage of OJ systems in educational environments and afterwards we briefly report the most relevant approaches for predicting student performance in the aforementioned context.

\subsection{OJ systems}\label{subsec:OJSystems}

The work by~\cite{hext1969}, who was the first to propose that academic computing assignments could be automatically graded, is considered the main precursor of current OJ systems. Nevertheless, their first formal definition was introduced by \cite{kurnia2001online} who described them as a computer system that automatically grades programming assignments and provides some type of feedback to the students. 

Regarding their practical use, the scientific literature comprises a large number of OJ proposals related, to a great extent, to academic institutions and educational environments. Some examples of such systems comprise the work by~\cite{carrasco2010aprendizaje} with the \textit{Javaluador} method for tasks in the Java programming language (it is described later in this paper), the \textit{URI} system by the Universidade Regional Integrada for developing and improving general coding skills~\cite{bezetal2014}, the \textit{Peking University Online Judge} (POJ) by~\cite{wen2005peking} tailored to C++ courses, the \textit{CourseMaker} one by the University of Nottingham for general programming tasks~\cite{Higgins2005}, the \textit{Youxue Online Judge} (YOJ)~\cite{sun2014yoj} also for improving coding skills inspired on exercises from different programming contests, and the \textit{Sphere Online Judge} (SPOJ) devised for E-Learning frameworks~\cite{kosowski2008spoj}, among others. 

Besides their use for educational purposes, OJ systems are also commonly considered in the context of coding competitions for solving algorithmic problems. Examples of such cases are the one used in the \textit{International Collegiate Programming Contest}~\cite{DEBOER2019104382} or the \textit{UVa} one considered in the Olympiads in Informatics~\cite{revilla2008competitive}.

\subsection{Estimating student performance}

The identification of struggling students in early course stages is deemed as a remarkably important topic in the education field as it suggests the instructor to provide additional resources to address the problem. In this sense, a large number of studies have assessed the influence of both extrinsic and intrinsic factors on the commented difficulties.

In relation to the extrinsic aspects, most of the existing literature resorts to the analysis of the socioeconomic position of the student or the marks obtained in previous courses\review{~\cite{alturki2021using}}. The reader is referred to the manuscript by~\cite{alturki2020predicting} for a thorough revision of these factors as it is out of the scope of this work. 

Regarding the intrinsic aspects---using information about the outcomes of the assignments carried out within a course---, the related literature comprises a large number of approaches since they typically yield considerably accurate predictions. Some representative examples include: the work by \cite{adnan2021}, which addresses this task in generic online learning platforms; that by \cite{cabral2019} on preventive failure detection in the context of the Moodle platform; the case of \cite{liao2019clicker} that estimates this information relying on information gathered from \emph{clicker} tests in peer-based instruction environments; and the approach by \cite{gray2019utilizing}, who use course attendance as a predictor of academic outcome for the academic year.

Focusing on the case of programming courses, it may be checked that the most basic, yet successful, approaches rely on hand-crafted heuristics neglecting the use of OJ systems. For instance, \emph{Error Quotient}~\cite{jadud2006} together with its refined version \emph{Repeated Error Density}~\cite{becker2016} perform this assessment by resorting to the syntax errors that occur during the compilation stage. The Watwin Scoring Algorithm~\cite{watson2013} works in a similar way, but penalises students based on the time required to fix each type of error compared to that of their peers. \cite{carter2015} devised a scoring mechanism that takes into account more complex interactions, such as debugging or modifying syntactically correct code. A last example is the one by \cite{tabanao2011} that identifies at-risk students by means of a linear regression approach based on compilation errors and other indicators. 

While the previous approaches are useful for addressing beginner-level programming courses, when tackling cases in which students are more familiar with this discipline, they become limited. In this context in which syntactic-level errors are less common than semantic ones, most approaches rely on the use of OJ systems and ML-based analysis techniques. Examples in the literature include the work by~\cite{castro2017} that proposes the use of a supervised classifier to predict final grades based on activity data, that of \cite{pereira2020} that studies the correlation between the different features from data related to the assignments of the students and the final grades with linear models, \cite{sym12040601} that addresses the problem as an exploratory factor analysis task, or the work by~\cite{azcona2019detecting} that combined data from an OJ with static information about the students---demographic information or academic marks obtained before enrolling in the course, among others---to predict their performance before each intermediate exam and, accordingly, suggest corrective actions with those who are likely to underperform. Note that, while successful, the main drawback of these proposals is the lack of interpretability of the derived models.

% We refer the reader to the work by \cite{hellas2018} for a more systematic review of approaches aiming at predicting student performance.

This work frames in the latter case of OJ systems for assessing coding tasks in programming courses. More precisely, our approach aims to predict the performance of the students out of the meta-information gathered from the submissions to the OJ for providing the corresponding feedback to both instructors and students. For that, we resort to both ML and MIL techniques for inferring these student profiles with the particular novelty of considering XAI approaches so that this feedback may be deemed interpretable. The next section thoroughly details the proposals.

%--------------------------------------------------------------------------
\section{Proposed methodology}
\label{sec:method}

%Predicting the student performance when addressing programming assignments and providing feedback accordingly represents a considerably wide statement that requires to be further particularised. In this regard, this work poses a set of research questions that our methodological proposal is will answer:
\review{This work poses the following research questions related to the prediction of student performance and feedback generation in the context of OJ assignments: }

% RQ1
\begin{hyp}[RQ\ref{hyp:first}] \label{hyp:first}
When should students start making submissions to the OJ system?
\end{hyp}

% RQ2
\begin{hyp}[RQ\ref{hyp:second}] \label{hyp:second}
How many submissions are reasonable for the success of the assignment?
\end{hyp}

% RQ3
\begin{hyp}[RQ\ref{hyp:third}] \label{hyp:third}
Which would be the risk groups?
\end{hyp}

% RQ4
\begin{hyp}[RQ\ref{hyp:fourth}] \label{hyp:fourth}
Are there any pieces of advice for the students to adequately address the assignments?
\end{hyp}

\begin{figure*}[!ht]
\centering  
  \includegraphics[width=.65\linewidth]{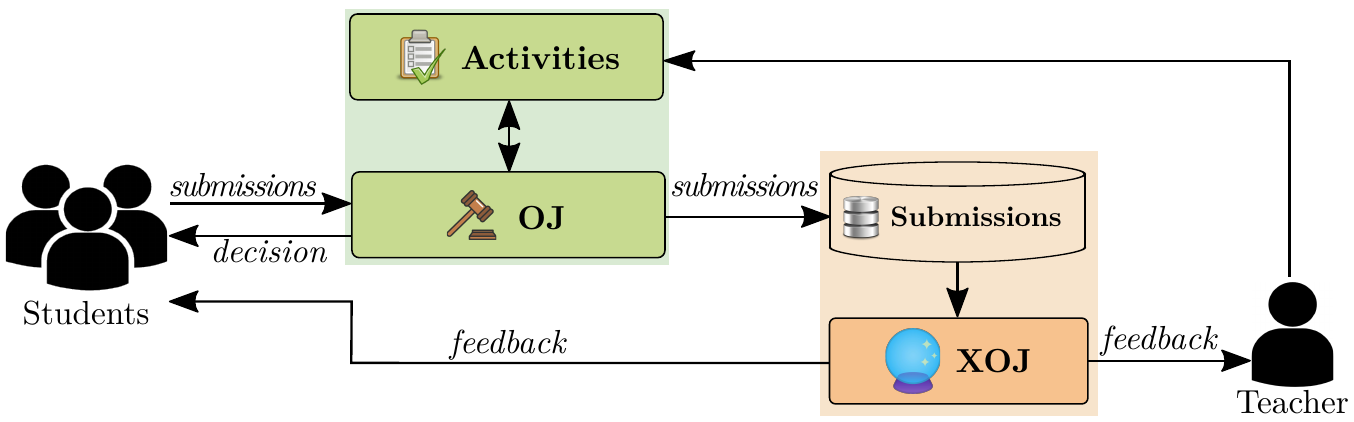}
  \caption{\review{Graphical representation of the scheme proposed.} %Initially, the teacher provides a set of activities and their respective corrections to the OJ; the students implement and submit these tasks to the system, which informs them about the correctness of the actual code; these submissions are then assessed by the XAI-based module---namely XOJ---devoted to providing feedback to both the students and the teacher about the development of the work. Note that the teacher may adjust the set of activities according to the aforementioned feedback.
  }
  \label{fig:scheme}
\end{figure*}

Figure~\ref{fig:scheme} graphically shows the scheme proposed to quantitatively address these questions, which comprises the following steps:

\begin{enumerate}
    \item The teacher defines the different assignments to be solved by the students and configures the OJ system accordingly.
    \item The students address the posed task and submit their implementations.
    \item The OJ evaluates these submissions and provides the students a correction mark exclusively based on the evaluation of the submitted programming codes.
    \item Concurrently, these submissions are processed by an additional module---XOJ in the scheme---that provides feedback to both the teacher---who may adapt the difficulty of the task---and the students---who may accordingly adjust their commitment to the task. Note that this element represents the core element of the work as it is meant to model the user behaviour considering a supervised learning framework. 
\end{enumerate}

%As commented, the EDM literature comprises a wide range of schemes capable of modelling and predicting student performance. In our case, we approach this task considering two learning-based schemes within a supervised framework: 
\review{We consider two learning-based schemes within a supervised framework for the XOJ: }
(i) a first one based on an MIL methodology that considers the introduced bag-of-instances concept to model the different submissions made by the students; and (ii) a second one that proposes an adaptation of the former case to standard ML algorithms and thus enables the use of XAI techniques so that human-understandable feedback may be derived out of the learning-based model.

The rest of the section further develops the concept of MIL frameworks, introduces the need for adapting these schemes to an ML-oriented task by means of the procedure proposed in the work, and finally presents the XAI scheme devised for answering the research questions posed as well as providing interpretable feedback to the users of the system.

\subsection{Multi-Instance Learning (MIL)} % approach
\label{sec:mil}

MIL represents a specific branch of supervised learning within the wider area of ML specifically devised to deal with incomplete knowledge of labels in corpora. This framework works on the basis of \textit{bags of elements}, \textit{i.e.}, collections of instances that, as a group, represent a certain element. In this regard, bags are labelled on a binary basis---either positive or negative---and the learning goal is to predict the class of unseen bags. \review{For a more in-depth introduction to the topic, the reader is referred to the work by~\cite{amores2013multiple}.}
%that provides a review of the different existing methods by elaborating a taxonomy of approaches.

\begin{figure}[!ht]
\centering  
  \includegraphics[width=.75\linewidth]{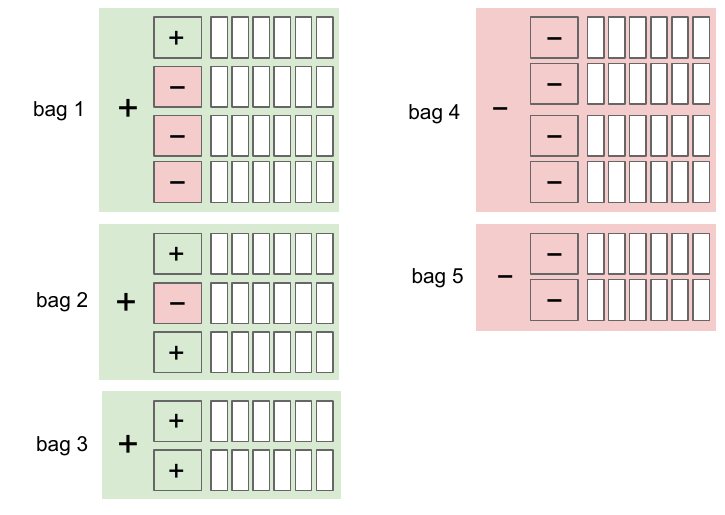}
  \caption{Toy example of the representation of bag of elements in MIL. The three bags on the left are labelled as positive since they contain, at least, one positive element; the two on the right are tagged as negative since they contain no positive instances.}
  \label{fig:MIL_bag_samples}
\end{figure}

During the training stage, these bags are assigned a class based on those of the individual instances within: a bag is labelled as positive if there is, at least, one positive instance in it, while it is considered negative if all instances are negative. Figure~\ref{fig:MIL_bag_samples} exemplifies this principle.

This MIL paradigm naturally fits the scenario posed in the work since each student may be modelled as a bag where the individual instances within represent the different submissions delivered to the OJ. \review{In line with previous works (\textit{e.g.},~\cite{ZafraRomeroVentura:HEDM:2010,ZafraVentura:ASOC:2012}), each bag is labelled as either positive or negative depending on whether the student succeeded in the assignment.}
%Note that, as previously discussed, this assumption is supported by other works in the literature, such as that by~\cite{ZafraRomeroVentura:HEDM:2010} or \cite{ZafraVentura:ASOC:2012}, that also consider MIL techniques for related EDM tasks.

%However, 
\review{As mentioned previously,} MIL frameworks generally work as \textit{black boxes}, \textit{i.e.}, a human user can neither know nor interpret the motivation of the model to perform a certain prediction. Hence, since explainability and interpretability represent a major concern in the more general ML field, the related literature comprises a large number of XAI works for learning-based schemes~\cite{arrieta2020explainable}. Nevertheless, these techniques are not suitable for MIL scenarios, being then necessary to previously adapt them to a general ML case. 

To perform such an adaptation we pose the following procedure: during the training stage, since all instances within an MIL bag are somehow represented by the class of the bag itself, we associate the individual instances after this general label and dismiss this MIL-based grouping; during the inference stage, every single instance is evaluated individually and the overall label is estimated by integrating these predictions. For this latter merging policy, this work proposes selecting the label with the maximum confidence score among the different learning-based models. It must be highlighted that initial experimentation also contemplated other statistical descriptors, being the commented maximum operator the one depicting the best overall performance.

\subsection{Explainable Artificial Intelligence for model interpretation}
\label{subsub:XAI}

%In a broad sense, 
\review{XAI techniques are typically divided into two different families}
%of approaches
~\cite{KENNY2021103459}: (i) \textit{transparency methods}, which represent the ones that directly convey the workings of the model; and (ii) \textit{post-hoc explanations}, which attempt to provide justifications about the reason why the model arrived at its predictions. This work frames on the latter case since, oppositely to transparency-based approaches, they avoid the need for individually adapting each learning-based model considered for the particular task at hand.

Within this \textit{post-hoc} framework, the two most common approaches rely on either performing permutations on the values of each predictor (independent variable) to assess their influence on the predictions (dependent variable); or creating an alternative and simple-to-explain linear model that mimics the behaviour of the one at issue~\cite{arrieta2020explainable}.

%Given that the latter family of approaches generally achieves more accurate results than the former one, we consider a specific approach within them. 
\review{We choose the latter family of approaches, as it generally achieves the most accurate results.}
More precisely, we resort to the so-called Shapley Additive Explanations (SHAP) introduced by~\cite{lundberg2017unified}, as it represents one of the most commonly considered techniques for XAI in the literature~\cite{fryer2021shapley}. SHAP is based on the concept of Shapley values, a well-known technique in the cooperative game theory field devised to measure the individual contribution of a player to the game~\cite{roth1988shapley}. The gist behind SHAP relies on calculating the Shapley values for each feature or predictor of the individual sample to be interpreted, representing each of these scores the impact on the prediction generated by the feature to which it is associated.

\begin{figure}[!ht]
% Juanra: diseño en https://docs.google.com/drawings/d/1j2jgRR8RpkXUGQhvyMzVTgk4BmL2UiOK9fWKLuz6YEQ/edit
\centering  
  \includegraphics[width=0.85\linewidth]{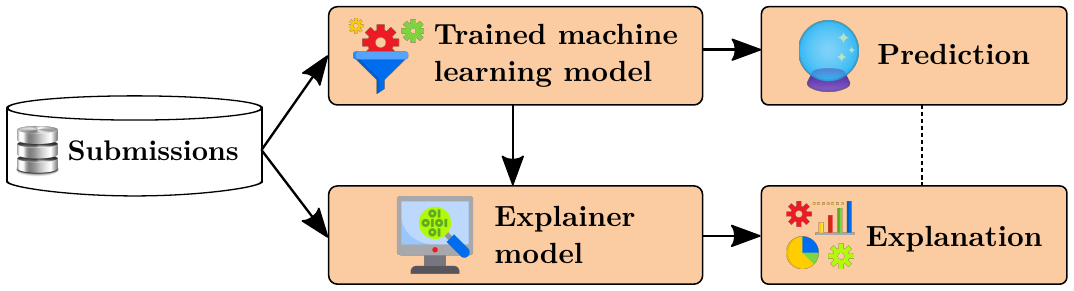}
  \caption{Conceptual description of the post-hoc XAI strategy considered \review{in which, for a given submission, the system both hypothesises on the possible success/failure of the student and provides the corresponding interpretable explanation. The dashed line denotes that the \textit{Explanation} block considers the \textit{Prediction} as a black box, \textit{i.e.}, it has no access to the internal states of the method.}}
  \label{fig:SHAP_scheme}
\end{figure}

On a practical basis, Fig.~\ref{fig:SHAP_scheme} conceptually shows the implemented pipeline for obtaining an interpretable prediction (feedback to be provided) for a query element (individual submission by a certain student) based on the SHAP method. Note that this process corresponds to the XOJ block depicted in the general scheme of the proposal (Fig.~\ref{fig:scheme}) \review{and matches that of a typical posthoc-oriented XAI framework.}

As shown in the scheme, a particular submission of a student---input data---is provided to the learning-based model; the prediction, along with the initial data, are then processed by the SHAP module that computes the corresponding Shapley values; the prediction, together with these values, are used for providing the corresponding feedback to both the student and the instructor.

%--------------------------------------------------------------------------
\section{Case of study}
\label{sec:study_case}

The case of study relates to the \textit{Programming Challenges} course of the senior year of the \textit{Degree in Computer Science} at the \textit{University of Alicante}. The evaluation in this course relies in two three-week-long assignments, denoted as A1 and A2 in this manuscript. \review{Note that the delivery periods are extended to compensate for the bank holidays that exist within the lapses of time stipulated for the different assignments}. These two assignments deal with optimisation strategies for algorithms and data structures and, more precisely, with the use of \textit{dynamic programming}---assignment A1---and \textit{branch-and-bound}---assignment A2---techniques~\cite{wirth1985algorithms}. The detailed description of these tasks is provided in \review{Appendix B}.
%\ref{app:problems}.

These assignments are developed autonomously by the students with a weekly supervision of the instructor during the lecturing sessions, in which specific aspects or doubts related to the tasks are discussed. Throughout the process, the students submit their implementations to an OJ that informs about the correctness of the submitted code. This OJ is the so-called \textit{Javaluador}~\cite{carrasco2010aprendizaje}, which was already introduced in Section~\ref{subsec:OJSystems} and is thoroughly described in Section~\ref{sec:javaluador} due to its relevance in this work.

The data used in this case of study was collected from all the actively developing the commented course---approximately, 80\% of the people enrolled in it---during three academic years (2019-2020, 2020-2021, and 2021-2022), being 2020-2021 a representative example of a remote-oriented scenario due to the COVID-19 pandemic situation. Table~\ref{tab:avg_subject} presents a description of this collection in terms of the number of students, amount of submissions to the OJ, and success rate.

% Note that the \textit{Submissions until success} column 

\begin{table*}[!ht]
    \caption{Description of the data collected for this study. For each academic year and assignment, the number of students enrolled, amount of submissions, and statistics about success/failure of the task are provided. 
    %\review{Note that \textit{Attempts} refers to the individual trials submitted by the students while \textit{Average submissions per student} refers to the overall result of the task.}
    \review{Note that the success/failure counts under the \textit{Attempts} column refer to the individual trials submitted by the students while those counts under the \textit{Average submissions per student} refer to the overall result of the task.  }
    }
    \label{tab:avg_subject}
    \centering
    \setlength{\tabcolsep}{5pt}
    \begin{adjustbox}{width = .95\textwidth, keepaspectratio}
        \begin{tabular}{cccccccccccc}
\toprule[1pt]
\multirow{3}{*}{Year} & \multirow{3}{*}{Assignment} & \multicolumn{3}{c}{\multirow{2}{*}{Students}} & \multicolumn{3}{c}{\multirow{2}{*}{\review{Attempts}}} & \multicolumn{2}{c}{\makecell{Average submissions\\per student}} & \multirow{1}{*}{\makecell{\review{Attempts}\\until success\\\review{(omits task failures)}}}\\
\cmidrule(rl){3-5} \cmidrule(rl){6-8} \cmidrule(rl){9-10} 
& & Total & Success & Failure & Total & Success & Failure & Success & Failure & \\
\cmidrule(lr){1-11}
\multirow{2}{*}{2019-20}
 & A1 & 21 & 13 & 8 & 317 &  7.9\% &  92.1\% & 13 & 18 & 12.3  \\
 & A2 & 18 & 11 & 7 & 316 &  7.6\% &  92.4\% & 24 &  8 & 22.2  \\
\cmidrule(lr){1-11}
\multirow{2}{*}{2020-21}
 & A1 & 37 & 31 & 6 & 626 & 43.3\% &	56.7\% & 17 & 15 &  8.7  \\
 & A2 & 33 & 27 & 6 & 452 & 25.4\% &	74.6\% & 14 & 14 &  9.2  \\
 \cmidrule(lr){1-11}
\multirow{2}{*}{2021-22}
 & A1 & 28 & 25 & 3 & 411 & 13.9\% &	86.1\% & 15 & 11 &  13.6  \\
 & A2 & 24 & 21 & 3 & 431 & 16.7\% &	83.3\% & 17 & 27 &  13.9  \\
\bottomrule[1pt]
        \end{tabular}
    \end{adjustbox}
\end{table*}

% \multicolumn{2}{p{3cm}}{Average submissions per student}

% This system, Javaluador, is also used in the final practical exam of the subject. Therefore, understanding and solving these problems correctly during the four-month period makes it easier to solve similar problems in the exam. The only difference is that in the practical exam with computer they have a maximum of 2h 30' (during the practicals they have up to three weeks) to solve one of the two new problems proposed, they have access to the code of the submissions made during the course but they do not have access to the Internet or any other electronic device.

% In this study, we will analyse the profile of each student through their different submissions with their results to Javaluador during the practices A1 and A2, and whether they have finally solved the problems successfully or not.
% \victor{mencionar que no se usan las notas finales del examen}

\subsection{Online Judge: Javaluador}
\label{sec:javaluador}

The \textit{Javaluador} OJ system~\cite{carrasco2010aprendizaje} comprises a collection of over 55 open-ended optimisation problems specially designed to be addressed with dynamic programming or branch-and-bound techniques. As suggested by its name, this OJ assesses tasks implemented in the \textit{Java} programming language.

The students enrolled in the course may use this bank of problems to practise and complement the theoretical contents explained during the lecture sessions since \textit{Javaluador} is entirely available throughout the academic year with no time restrictions. Note that this statement does not apply to the aforementioned A1 and A2 evaluation assignments since the system restricts the submissions of these two tasks to their respective evaluation periods.

\textit{Javaluador} evaluates the efficiency and correctness of the submitted programs. For this purpose, it imposes two restrictions: (i) the provided implementation must not exceed a time limit in solving the entire battery of tests designed for the problem, which is fixed to 10 seconds; and (ii) each test must not exceed a certain amount of memory, which is set to 100~MB.

Once the system has assessed the submission, the OJ provides the student with a decision about the success of the task in one of a set of 6 possibilities:\footnote{The reader is referred to Fig.~\ref{fig:scheme} in which the OJ provides a \textit{decision} to the student solely based on the correctness of the provided code.} (i) \textit{success} if the submission passed all tests considering the aforementioned time and memory constraints; (ii) \textit{test error} if only a subset of the tests were successfully passed; (iii) \textit{compilation error} if the program did not compile correctly---the actual error is also provided to the student---; (iv) \textit{time error} if the time limit was exceeded; (v) \textit{memory error} when the execution memory was exceeded; and (vi) \textit{function error} when the student included functions or libraries discouraged from their use due to security and/or educational reasons.

\subsubsection{Variables}
\label{sec:variables}

\review{In order to provide feedback about the general performance of the student within the actual assignment, our proposed XOJ makes use of the set of variables defined next. Note that these variables must be available while the student is developing the assignment. Hence, they must not refer to the overall correction of the assignment---they may include, however, the correctness of the current submission---but 
instead they contain information that can be provided by the OJ.}
%This section defines the different variables considered in this work for inferring the corresponding feedback related to the submissions. 
%Note that, since these observations must be provided while the student is developing the assignment, we must define a set of descriptors that do not refer to its overall correctness---it may include, however, the correctness of the current submission---and that the OJ may provide. In this regard, we propose the following set of variables:

\begin{itemize}
    \item \textit{Days to deadline}: Number of remaining days from the current submission until the deadline of the assignment. This variable measures the closeness to the deadline date to find out whether the student has delivered early, has had several submission periods over the duration of the problem, or has delivered very close to the end date. This descriptor is a real value in which the integer and decimal parts respectively represent the complete and the portion of the remaining days to the deadline.
    
    \item \textit{First submission \review{--} days to deadline}: This descriptor, which follows the same format as the \textit{Days to deadline} variable, represents the number of remaining days to the deadline when the student submitted the first attempt of the assignment. \review{Note that this value is set when a student performs the first submission to an assignment and remains constant for the subsequent submissions.}
    %Note that this value is set following the commented procedure for each student and assignment and remains constant for the rest of the submissions.
    
    \item \textit{Submissions to date}: Number of previous submissions made so far including the current one. This variable somehow relates to the insistence of the student in solving the tasks or, if already done, the willingness to refine them.
    
    \item \textit{Submission days to date}: Number of different days on which a certain student has submitted different attempts to solve the assignment. % With this variable we can measure whether the student has made successful or unsuccessful submissions over a period of time or over several periods of time.
    
    \item \textit{Assignment}: It represents whether the submission relates to assignment 1 (A1)---corresponding to the \textit{dynamic programming} task---or to assignment 2 (A2)---corresponding to the \textit{branch-and-bound} approach.
    
    \item \textit{Success}: \review{It represents whether the submission belongs to a student that passed or failed the assignment.}
    %Represents whether the result of the current submission belongs to a successful or failed historic of attempts.
    
\end{itemize}

It must be noted that, due to the different ranges in which these variables may span, a standard normalisation process is applied to avoid any possible biases in the learning-based schemes.

% Section~\ref{sec:study_case}
% Finally, showsFigure~\ref{fig:hist_submissions} shows the number of submissions made relative to the number of days remaining to complete the problem. We can see how successful submissions are made regularly throughout the duration of the problem. The total submissions increase as the deadline approaches except for intervals including holidays or weekends.

% \begin{figure}[!ht]
% \centering  
%   \includegraphics[width=.75\linewidth]{histogram_Days_to_deadline_all.pdf}
%   \caption{Histogram of the number of submissions relative to the number of days to deadline.}
%   \label{fig:hist_submissions}
% \end{figure}

% ---------------------------------------------------------------------------------------
\section{Experimental setup}
\label{sec:setup}

This section details the experimental arrangement considered in the work in terms of evaluation metrics, validation policies, and learning-based schemes. Regarding software tools, we have performed this analysis using a variety of open-source tools: Python was used as the base programming language as well as the libraries \textit{scikit-learn} (v0.24.0), \textit{xgboost} (v0.90), and \textit{catboost} (0.24.0) to implement ML algorithms, \textit{MIL} (v1.05) to implement Multi-Instance Learning approaches, and \textit{SHAP} (v0.37)~\cite{SHAP} to analyse XAI-ML aspects.

\subsection{Metrics}
\label{sec:metrics}

To assess the goodness of the proposal, we consider the Area Under the Receiver Operating Characteristics Curve, which is denoted as AUC in the work. This metric, which is typically considered in both classification and regression tasks involving threshold settings, penalises larger differences between the real and the predicted value~\cite{DudaHartStork01}. Mathematically, AUC is defined as:

\review{
\begin{equation}
    \mbox{AUC}(\hat{f}) = \frac{1}{|\mathcal{Y}^0|\cdot|\mathcal{Y}^1|}\sum_{t_0\in \mathcal{Y}^0} \sum_{t_1\in \mathcal{Y}^1} \mathbf{1} [\hat{f}(t_0) < \hat{f}(t_1)]
    \label{eq:AUC}
\end{equation}
\noindent where $\mathcal{Y}^{0},\mathcal{Y}^{1}\in\mathbb{N}^{n}$ denote the $n$-sized spaces that respectively represent the set of prototypes labelled with 0 (failure) or 1 (success), $\hat{f}:\mathbb{N}^{n}\rightarrow\left[0,1\right]$ represents the estimator obtained by the considered ML or MIL method, and $\mathbf{1} \left[\cdot\right]\rightarrow\left\{0,1\right\}$ denotes an indicator function that returns 1 if the condition in the argument is fulfilled and 0 otherwise. Note that the size of the vector matches the number of features used by the model, \textit{i.e.}, $n=5$ descriptors (see Section~\ref{sec:variables} for the detailed description of these elements).}

% Justificación del uso de la métrica AUC y la validación cruzada de 10.
A commonly considered procedure to avoid biased results in the evaluation of schemes involving learning-based methods is the so-called stratified cross-validation~\cite{Mitchell97}. This strategy divides the available data into a fixed number of partitions---usually denoted as folds---and uses all sets except one to train the learning-based method and the remaining set to evaluate its performance. This process is repeated as many times as the established number of folds, which results in that all data has been used both for training and evaluation. \review{Eventually, the overall performance of the scheme is computed as the average of the individual scores obtained in the different folds.}

\begin{figure}[!ht]
     \centering
     \begin{subfigure}[b]{.5\textwidth}
          % source: https://docs.google.com/drawings/d/1qgsmcF7opvSB7NmO6trjt08cCfAx7U-hjFO_aKgdwIA/edit
          \centering  
          \includegraphics[width=.9\linewidth]{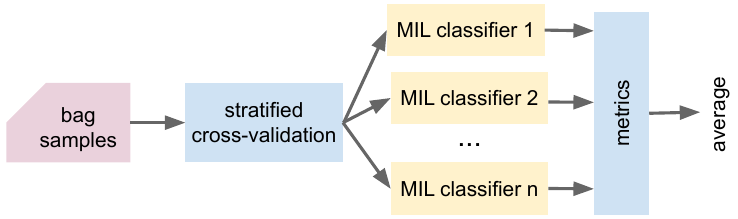}
          \caption{Stratified cross-validation scheme for MIL problems. The partitioning is carried out at the bag level, hence being each of these sets in only one partition.}
          \label{fig:MIL_CV}
     \end{subfigure}\\
     \begin{subfigure}[b]{.5\textwidth}
          % source: https://docs.google.com/drawings/d/1-0ZroglqPSAu9cgUB4bThKJM-QGmpVJRdaegmObWacI/edit
          \centering  
          \includegraphics[width=.9\linewidth]{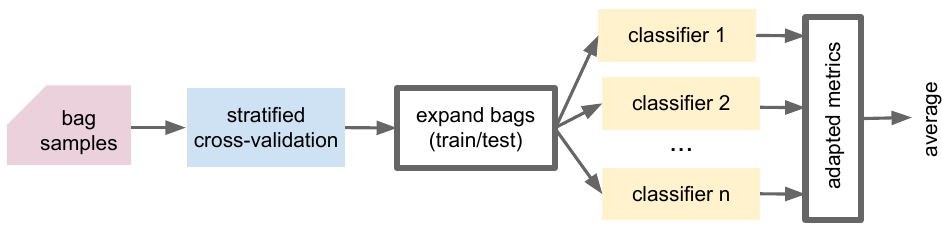}
          \caption{Stratified cross-validation scheme for ML problems. In this case, while there is no longer a concept of bag, all instances that originally belonged to the same bag must be in the same partition to avoid any possible bias in the results.}
          \label{fig:ML_CV}
     \end{subfigure}
        \caption{Stratified cross-validation schemes for the MIL (top) and ML (bottom) approaches.}
        \label{fig:cross_validation}
\end{figure}

In the particular context of this work, it must be noted that, when performing the partitioning process, all elements within a bag must always be part of the same partition. While this point may be obvious in MIL representations, when considering ML strategies in which the concept of bag disappears, all elements that were formerly part of the same bag must also be in the same partition. Otherwise, we would induce some bias in the training and evaluation of the models. Figure~\ref{fig:cross_validation} graphically shows these ideas.

Finally, in this work, we resort to a value of 10 folds for this cross-validation approach as it represents a typical value in the related literature.

%----------------------------------------------------
\subsection{Learning algorithms compared}
\label{sec:algorithms}

% https://github.com/rosasalberto/mil

To validate our proposal, we have considered a representative collection of learning-based methods from both the MIL and ML families of techniques. We now present the different algorithms studied, which are also summarised in Table~\ref{tab:algorithms} together with the parameters used in the experimentation.

Regarding MIL methods, we have assessed the following approaches: 

\begin{itemize}
    \item \textit{Axis Parallel Rectangle} (APR)~\cite{dietterich1997solving}: Iterative discrimination algorithm that finds an axis-parallel hyper-rectangle in the feature space to represent the target concept, \textit{i.e.}, all the instances from positive bags.
    
    \item \textit{Citation-$k$NN}~\cite{wang2000solving}: Adaptation of the well-known $k$-Nearest Neighbour ($k$NN) classifier~\cite{NNCover1967} to MIL frameworks.
    
    \item \textit{Diverse Density} (DD)~\cite{maron1998framework}: Method that seeks to find a concept point in the feature space close to, at least, one instance from every positive bag.
    
    \item \textit{Expectation-Maximisation Diverse Density} (EM-DD)~\cite{zhang2001dd}: Extension of the DD strategy considering an Expectation-Maximisation method.
    
    \item \textit{Multiple-Instance Learning via Embedded Instance Selection} (MILES)~\cite{chen2006miles}: Approach that maps each bag into a feature space defined by the instances in the training bags via an instance similarity measure.

    \item \textit{Deep Attention MIL}~\cite{ilse2018attention}: Strategy that estimates the contribution of each instance to the overall bag label by framing the problem as learning a Bernoulli distribution in which their probabilities are parameterised by both neural networks and attention mechanisms.
\end{itemize}

In addition to this, we have also assessed the use of preprocessing techniques from the literature known as \textit{bag representations} that allow tackling MIL-based tasks with binary ML classification algorithms by mapping these bag representations into instance-based feature vectors. The representative set of techniques considered comprises the introduced MILES method~\cite{chen2006miles}, the so-called Discriminative approach~\cite{wu2018multi} and the use of basic statistical methods based on the mean aggregation of the individual instances within the bags.

Finally, we have also studied the use of standard ML methods in the proposed task. Note that, since these techniques are not directly applicable to MIL-oriented problems, we resort to the policy proposed in Section~\ref{sec:mil} that performs an MIL-to-ML adaptation. In these terms, we consider the following collection of ML algorithms: 

\begin{itemize}
    \item \textit{Naive Bayes} (NB)~\cite{Webb2010}: Classifier based on the Maximum A Posteriori criterion which assumes that the different features in the instances are statistically independent.
    
    \item \textit{Logistic regression} (LogReg)~\cite{menard2002applied}: Binary classifier based on the linear combination of the predictors.
    
    \item \textit{$k$-Nearest Neighbours} ($k$NN)~\cite{NNCover1967}: Non-parametric method that considers a certain dissimilarity metric for classification.
    
    \item \textit{Decision Tree} (DT)~\cite{breiman2017classification}: Classifier based on the principle of information gain that produces a tree-like model to perform the process.
    
    \item \textit{Random Forest} (RF)~\cite{breiman2001random}: Ensemble method based on the construction of multiple DT whose individual predictions are combined based on a specific criterion, most commonly the mode of the individually estimated classes. 
    
    \item \textit{Adaptive Boosting} (AdaBoost)~\cite{freund1997decision}: Meta-classifier based on the combination of a different number of classifiers. Oppositely to RF, the individual methods are not required to be based on a certain type of classifier. In addition to this, we also consider the \textit{eXtreme Gradient Boosting} (XGBoost)~\cite{XGBoostChenG16},  
    \textit{Categorical Boosting} (CatBoost)~\cite{dorogush2018catboost}, and \textit{Light Gradient Boosting Machine} (LGBM)~\cite{ke2017lightgbm} methods as they are based on the same principle of RF and are deemed as considerably competitive in the related literature.
    
    \item \textit{Support Vector Machine} (SVM)~\cite{Cortes1995svm}: Binary classifier based on the estimation of a hyper-plane to split the two categories involved in the problem.
    
    \item  \textit{Multilayer Perceptron} (MLP)~\cite{hinton1990connectionist}: Classifier based on a feedforward artificial neural network comprising multiple layers of perceptrons.
\end{itemize}

\begin{table*}[!h]
\caption{Summary of the different learning-based algorithms considered along with their identifiers and the evaluated parameters.}
\label{tab:algorithms}
\centering
\begin{adjustbox}{width = 0.9\textwidth, keepaspectratio}
\begin{tabular}{clll}
\toprule[1pt]
\textbf{Type} & \textbf{Algorithm} & \textbf{Identifier} & \textbf{Parameters evaluated} \\
\cmidrule(lr){1-4}
    % --  & Baseline  &  &  \\
%  \midrule
    \multirow{5}{*}{\rotatebox[origin=l]{90}{MIL}} 
        & Axis Parallel Rectangle \cite{dietterich1997solving}         & APR       & thresh.=0.5, $\epsilon$=0.05, step=1 \\
        & Citation-$k$NN~\cite{wang2000solving}                          &           & references=1; citations=3 \\
        & Expectation-maximization with Diverse Density~\cite{zhang2001dd} & EM-DD & scale=1, epochs=10, thresh.=0.5 \\
        & MIL via Embedded Instance Selection \cite{chen2006miles}     & MILES     & $\sigma = 4.5$, $C=0.5$  \\
        & Deep Attention MIL \cite{ilse2018attention}                  &   &  gated=False, thresh.=0.4 \\
\cmidrule(lr){1-4}

    \multirow{11}{*}{\rotatebox[origin=l]{90}{ML}} 
        & Naive Bayes~\cite{Webb2010}                              & NB       & Laplace smoothing = 1.0  \\
        & Logistic regression~\cite{menard2002applied}             & LogReg   & Stopping criteria tolerance = 1e-4  \\
        & $k$-Nearest Neighbours~\cite{NNCover1967}                  & $k$NN      & k=\{1, 3, 5, 7, 9, 11\}  \\
        & Decision tree~\cite{breiman2017classification}           & DT       & Split criteria = Gini impurity  \\
        & Random Forest~\cite{breiman2001random}                   & RaF      & est.=\{50, 100, 200, 300, 400, 500\}  \\
        & eXtreme Gradient Boosting~\cite{XGBoostChenG16}          & XGBoost  & booster=gbtree, $\eta=0.3$, max depth=6  \\
        & Categorical Boosting~\cite{dorogush2018catboost}         & CatBoost & Additive smooth = 1, iterations = 5  \\
        & Light Gradient Boosting Machine~\cite{ke2017lightgbm}    & LGBM     & booster=DT, num. leaves = 31 \\

        & Adaptive Boosting~\cite{freund1997decision}              & AdaBoost & estimators=50  \\
        & Support vector machine~\cite{Cortes1995svm}              & SVM      & kernel=\{linear, radial\}    \\
        & Multilayer Perceptron~\cite{hinton1990connectionist}     & MLP      & 1 hidden layer with 100 neurons  \\
\cmidrule(lr){1-4}
    \multirow{5}{*}{\rotatebox[origin=l]{90}{Bag repr.}}
        & Mean + RaF                                        &        & est.=\{50, 100, 200, 300, 400, 500\}  \\
        & Mean + LogReg                                     &        & Stopping criteria tolerance = 1e-4     \\
        & Mean + SVM                                        &        & kernel=\{linear, radial\}           \\
        & MILES + SVM                                       &        & kernel=\{linear, radial\}           \\
        & Discriminative \cite{wu2018multi} + SVM          &        & kernel=\{linear, radial\}           \\
\bottomrule[1pt]
\end{tabular}
\end{adjustbox}
\end{table*}

%----------------------------------------------------------------------
\section{Results}\label{sec:results}

This section presents the results obtained with the methodology proposed in Section~\ref{sec:method}
considering the case of study described in Section~\ref{sec:study_case}. More precisely, the different learning-based methods are initially compared in terms of the reported classification figures; after that, the capabilities of the XAI scheme proposed are examined, both in relation to the feedback provided as well as the recognition rate achieved; finally, a cohort study is performed to further provide insights about the capabilities of the proposal.

\subsection{Classification results}

\begin{figure}[!h]
\centering  
  \includegraphics[width=1.0\linewidth]{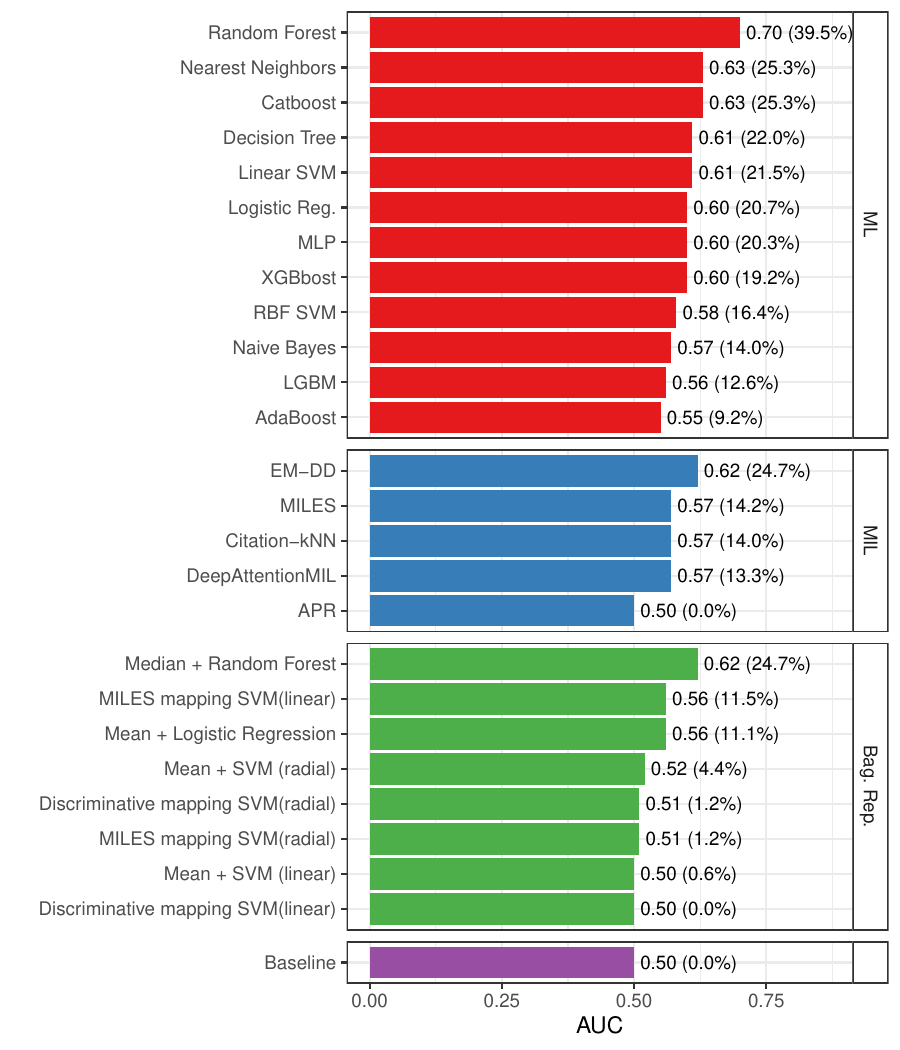} % Por bloques desde R, Juanra
  \caption{Average results for the different classification schemes in terms of the AUC metric. Relative improvement with respect to the baseline case is denoted in parentheses. For a better comprehension, results are sorted in decreasing AUC score and grouped according to learning-based family.}
  \label{fig:classification_results}
\end{figure}

Figure~\ref{fig:classification_results} shows the results obtained for the student performance prediction task---either success or failure---in terms of the AUC metric for the classification schemes and the experimental setup considered. For reference purposes, we include a \textit{Baseline} case in which the prediction method always outputs the majority class in the corpus. In all cases, these values represent the average of the 10 folds in the cross-validation scheme, being the detailed figures obtained in each fold of the cross-validation assessment strategy available in \review{Appendix A}.
%\ref{app:auc_ml_mil}.

A first remark that may be observed is that most of the elements in the different families outperform the baseline considered. Attending to the relative improvements, the baseline is enhanced up to 39.5\%, which means an improvement of 0.20 in absolute points.

Focusing on the ML family, it may be checked that all configurations within this group improve the baseline in a range between 9.2\% and 39.5\% of relative improvement. Note that the RF strategy yields the best overall result as it achieves a 0.70 of AUC score.

Regarding the MIL family, most approaches within this family also improve the base case in a range between 13.3\% and 24.7\% of relative boost. The sole exception to this is the APR which ties in performance with the baseline. On this note, the EM-DD yields the best classification rate with a 0.62 of AUC score.

In relation to the bag representations, most of the schemes also improve the base case of selecting the majority class. However, five out of the eight evaluated models show a marginal, or even null, increase in the performance---relative improvement between 0\% and 4.4\%. Within this group, the Median+Random Forest strategy obtains the best classification rate as it yields a 0.62 in the AUC score (relative improvement of 24.7\%).

Attending to this, the RF scheme with the proposed MIL-to-ML mapping proves to be the best performing strategy of the different schemes and methods evaluated.

\subsubsection{Significance tests}

\begin{figure*}[!h]
\centering
  \includegraphics[width=.75\linewidth]{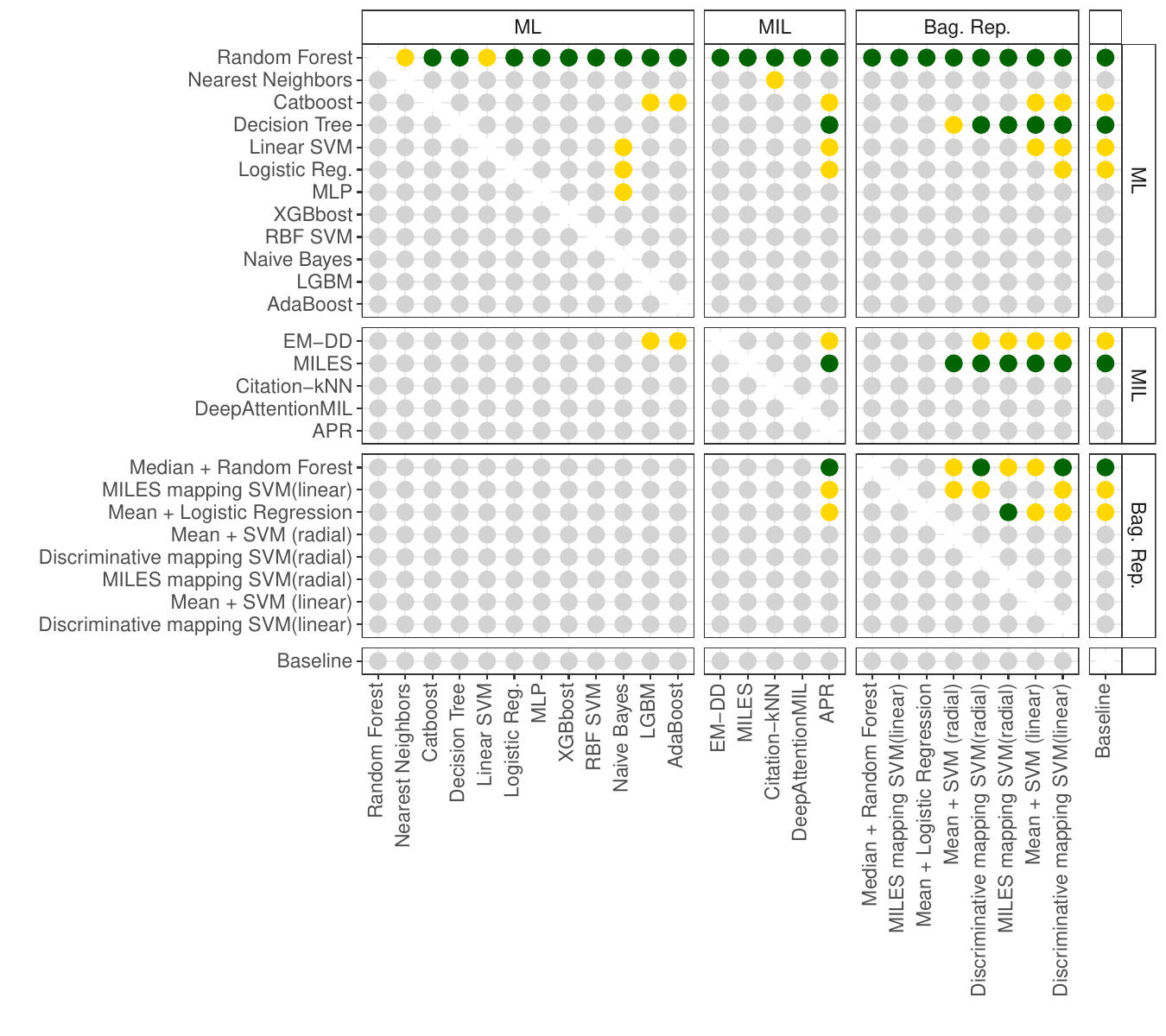} % Juanra: Hecho con R por bloques
  \caption{Wilcoxon signed-rank test of the pairwise comparison in terms of performance of the considered classification algorithms. Yellow and green colours indicate that the method in the row significantly improves that of the column with confidence values of 90\% and 95\%, respectively.}
  \label{fig:wilcoxon}
\end{figure*}

To further extend the previous analysis and provide a solid set of conclusions, we now perform a statistical evaluation of the results obtained. For that, we resort to the Wilcoxon signed-rank test~\cite{wilcoxon1945individual} and carry out a pairwise comparison of the different classifiers in terms of the performance considering the individual AUC figures for each fold and classifier. Figure~\ref{fig:wilcoxon} graphically provides the results obtained when considering the statistical significance levels of 90\% and 95\%, respectively depicted in yellow and green colours.

Attending to these results, a first remark to point out is that, out of each family of methods considered, only a subset of them statistically outperform the baseline considered. In general, these methods are the ones with a remarkable performance improvement in the respective families: RF, Catboost, DT, Linear SVM, and LogReg from the ML case; EM-DD and MILES from the MIL scenario; and Median + Random Forest, MILES + Linear SVM, and Mean + Logistic Regression from the bag representation. A particular exception to this assertion is the case of $k$NN from the ML family that, despite achieving a remarkable enhancement, there is no evidence of statistical improvement, most likely due to inconsistent behaviours in the different folds. This claim about the inconsistency of $k$NN may be checked in \review{Appendix A}, in which the values of the different folds are provided.
% \ref{app:auc_ml_mil}

In general terms, the RF method from the ML family stands as the most competitive scheme among the ones assessed since it statistically outperforms the rest of techniques. Note that this improvement is observed with a confidence value of 95\% in all scenarios except for two cases in which this score is lowered to 90\%: the $k$NN and the Linear SVM, also from the ML family of approaches. Due to this, we consider this RF method as the learning-based scheme for our proposal throughout the rest of the experimentation.

\subsection{XAI results}

The previous comparison posed the RF classifier as that which achieved the most significant and consistent improvement against both the baseline and the rest of alternatives. However, as aforementioned, the results provided by this scheme are not directly interpretable by a human, at least in a straightforward manner.\footnote{Note that, while tree-based classifiers are generally deemed as interpretable, when used as part of ensemble schemes---such as forest structures---this explainability characteristic is hindered.}

\begin{figure}[!h]
\centering  
  \includegraphics[width=1.0\linewidth]{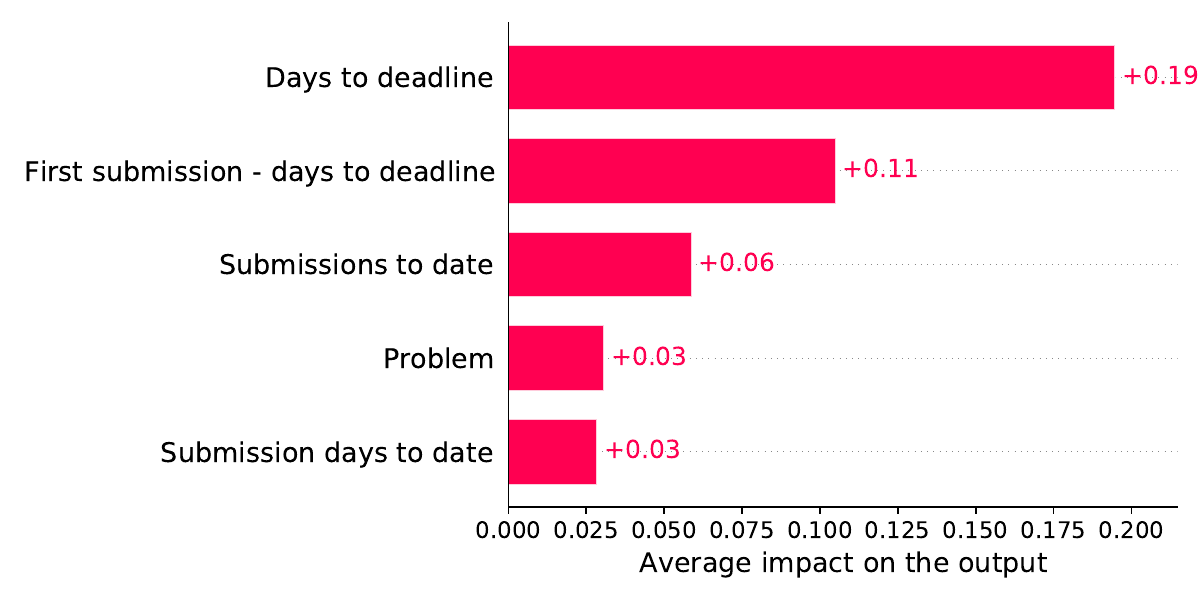}
  \caption{SHAP analysis of the input variables with respect to the classification rate for the RF method. Features are ranked from highest to lowest in terms of the obtained Shapley values.}
  \label{fig:general_shap}
\end{figure}

In this regard, we now perform the XAI analysis from our proposal (cf. Section~\ref{sec:method}) to state the influence in the outcome of the prediction method of each of the variables posed in Section~\ref{sec:variables}. For that, we consider the SHAP method introduced in Section~\ref{subsub:XAI} that estimates the impact of each independent variable on the success of the problem in terms of the Shapley values. Figure~\ref{fig:general_shap} shows the results of this study.

% Juanra: Impacto de los predictores sobre la superación del problema

As it may be checked, the \textit{Days to deadline} and \textit{First submission \review{--} days to deadline} stand out as the most relevant variables as they retrieve the highest Shapley values---roughly, their scores double those of the rest of the parameters. The \textit{Submissions to date} and \textit{Submission days to date} variables also depict some influence on the result, while the \textit{Problem} one has a marginal impact on the performance.

\begin{figure*}[!h]
\centering  
  \includegraphics[width=.6\linewidth]{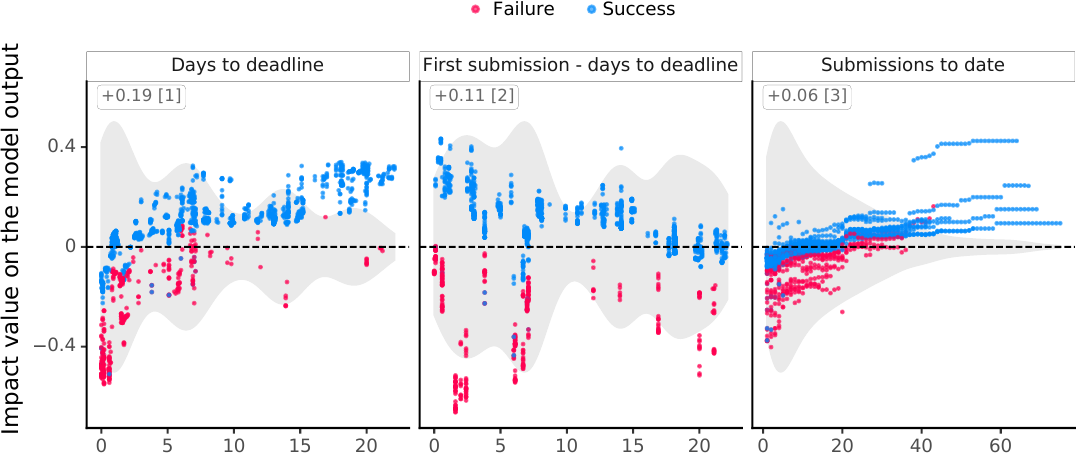}
  \caption{Individual impact on the output model against the three variables with the highest Shapley values of the SHAP analysis---\textit{Days to deadline}, \textit{First submission \review{--} days to deadline}, and \textit{Submissions to date}. The density of samples in each area is shown as a violin plot in the background. The average impact together with the position of the ranking are indicated at the top-left of each particular case. Ground-truth labels are provided using the colour of the point---red and blue for failing and passing the task, respectively.}
  \label{fig:individual_shap}
\end{figure*}

To further understand the influence of these variables, we now present an additional analysis of these variables. In this regard, Fig.~\ref{fig:individual_shap} graphically shows the individual scores obtained by the different variables in an isolated manner---assuming complete statistical independence among them---for each of the submissions in the case of study.

% Juanra: Desglose del impacto por individuo de los predictores sobre las variables de salida

% If we consider the variable \textit{submission days to date} values 2 and 3 are the best considered with positive impact while the rest are indistinct having positive and negative cases. It makes sense as the subject is taught as these problems are three weeks long with guided sessions once a week, which is when most of the students are engaged in handing them in %%% ESTO NO APARECE EN LA GRAFICA, POR LO QUE LO QUITO

Attending to these graphs, it may be observed that when tasks are submitted close to the deadline---\textit{Days to deadline} $< 7$ in the left graph---the system considers them as negative (vertical axis), which remarkably matches the ground-truth labels provided. Regarding the \textit{First submission \review{--} days to deadline} variable---middle graph---, there is no clear trend that relates the value to the success of the task since the results show cases of both students with early submissions who eventually fail and vice versa. Finally, the \textit{Submissions to date}---right graph---indicates that students submitting more than 40 attempts generally succeed in adequately solving the task. Note that this case does make sense as it corresponds to students who regularly use the system to develop the assignment.

\subsection{Cohort study}
\label{subsub:cohort}

The presented results in the previous sections draw general conclusions related to the relevance of the predictive variables in the output. Hence, to further extract additional insights about the reach of our proposal, this section develops a Cohort analysis of the scheme. More precisely, we consider the individual submissions by the students and divide them into different groups with related values of the variables---namely cohorts---to study their particular behaviour in the overall success of the task. Note that this grouping strategy is expected to provide some commonalities among the different student profiles based on the latent information of the submissions delivered.

\begin{figure}[!h]
\centering  
  \includegraphics[width=1.0\linewidth]{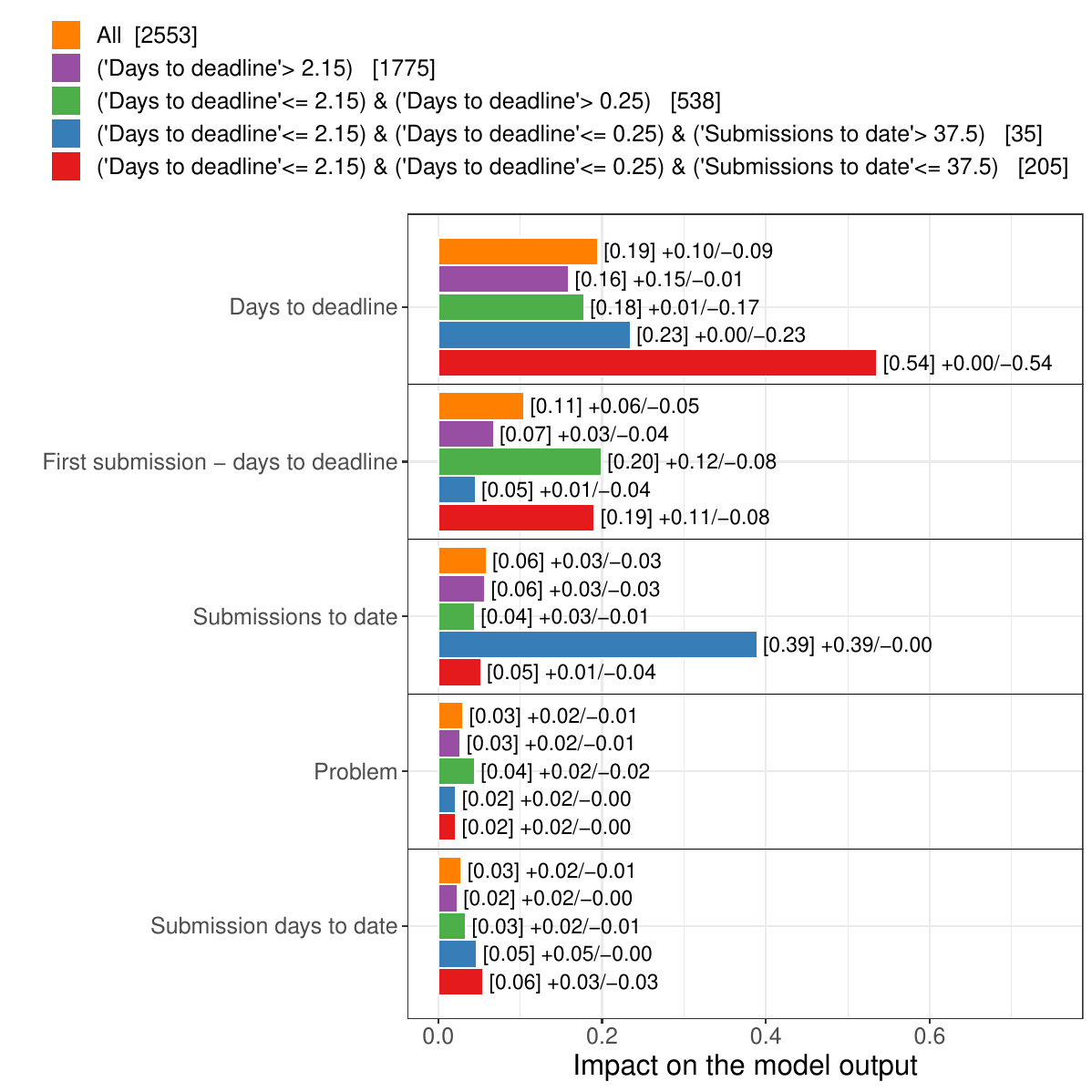} % Actualizado por Juanra
  \caption{Analysis of the influence of the posed variables in the XAI result considering a Cohort analysis with four user groups. \review{The symbols + and - detail the influence on the positive and negative classes, respectively.} Legend indicates the XAI-based rule and the number of students---in square brackets---included in the different cohorts. Label \textit{All} denotes the case when considering a single group with all students.}
  \label{fig:cohort_free_bar}
\end{figure}

\review{In order to automatically obtain these cohorts, we resort to one of the ML approaches previously introduced: the \textit{Decision Tree} (DT) model. Based on a defined information-gain criterion, this technique derives a set of interpretable rules that partitions the feature space of the collection of data at hand with the goal of maximising the classification rate. In this regard, we resort to the Gini impurity criterion to iteratively divide the initial set of samples into subsets until a maximum is reached---user parameter---as it constitutes one of the most common criteria used in DT (see Table~\ref{tab:algorithms})}. Figure~\ref{fig:cohort_free_bar} shows the average impact of predictors when fixing a maximum of 4 cohorts. It must be highlighted that this parameter was selected as it represents a trade-off between a general view---selecting a single group---and a totally particular view---choosing as many cohorts as individuals in the population.

Attending to the results obtained, the \review{\textit{`Days to deadline'$>$2.15} group}---namely Cohort A for its posterior analysis---proves to be \review{mainly influenced by the \textit{Days to deadline} variable as it depicts the highest impact---score of $0.16$---among all the descriptors.}

Regarding the \textit{`Days to deadline'$<$=2.15 \& `Days to deadline'$>$0.25} group---denoted as Cohort B for the latter analysis---, \review{it may be noted that it is mainly influenced by the \textit{Days to deadline} and \textit{Submissions to date} variables. These two descriptors respectively depict impact scores $0.18$ and $0.20$ whereas the rest of them achieve very low influence values (\textit{i.e.}, figures lower than $0.05$).}

In the case of the \textit{`Days to deadline'$<$=2.15 \& `Days to deadline'$<$=0.25 \& `Submissions to date'$>$37.5} group---namely Cohort C---it may be observed that it is highly influenced \review{by} the \textit{Submissions to date} variable, mostly in positive terms---score of $0.39$---with no remarkable influence \review{from} the rest of the features.

When examining the last cohort posed---\textit{`Days to deadline'$<$=2.15 \& `Days to deadline'$<$=0.25 \& `Submissions to date'$<$=37.5}, denoted as Cohort D---, it may be observed a score of $0.54$ on the \textit{Days to deadline} variable, \review{hence depicting a clear impact on this cohort.} 
% hence depicting a clear negative impact on this feature. 
While this group is also influenced by the \textit{First submission \review{--} days to deadline} variable, its overall impact is not as sharp as in the former case.

In addition to the individual analysis of the cohort scores obtained, we now examine the statistical significance of these results. For that, we consider the Wilcoxon Rank-Sum Test~\cite{wilcoxon1945individual} to compare the dependent variable---the performance of the scheme---against the influence score depicted by each cohort. In this case, the results considering a significance value of $\rho<0.05$ prove that the only significant influence is the one described for Cohort D---significance value of $\rho_{D}=0.0007$---while the results for Cohort B---significance of $\rho_{B}=0.1922$---and Cohort C---significance of $\rho_{C}=0.5725$---do not show any statistical differences among the distributions.

% -------------------------------------------------------------------------
\section{Discussion}
\label{sec:discussion}

Having presented the results of the different experiments posed and their corresponding analyses, this section provides a general summary and overall discussion of the main insights obtained together with the devised answers to the Research Question formulated.

As presented, the proposed methodology aims to identify student profiles---mainly, those likely to fail the proposed assignments---in the context of a programming-related course using OJ systems for the autonomous correction of the tasks. After gathering a collection of representative data comprising different tasks and academic years for this case of study, we proposed a set of hand-crafted descriptors by computing basic statistics directly retrievable from the OJ, such as the number of submissions, days to deadline, or assignment considered. Based on that, we studied different approximations to automatically model the user behaviour:

\begin{itemize}
    \item As a first approach, due to its reported success in EDM tasks, we have considered the use of MIL techniques for performing such an approximation. This case achieved a maximum score of 0.62 in the AUC figure of merit used in the work.
    \item Methods based on the so-called bag representations achieved the same maximum success rate of 0.62 in terms of the AUC metric.
    \item Finally, by performing the proposed method for adapting MIL representations to classic ML frameworks, we were able to boost this classification rate up to a value of 0.70 in the AUC figure of merit, which statistically outperforms all other schemes considered.
\end{itemize}

After obtaining the optimal classification strategy, we considered the use of XAI to obtain an interpretable model capable of relating the aforementioned descriptors with the actual outcome of the model. In this regard, factors such as the \textit{Number of days from the current submission to the deadline of the assignment} or the \textit{Time lapse from the first submission to the deadline of the assignment} proved to be highly influential while others such as the \textit{Task identifier} marginally affected the outcome of the model. This point suggests that the actual success of the tasks mainly relates to the individual aspects of the student---\textit{e.g.}, attitude and work compromise---rather than to the inherent difficulty of the assignment.

Finally, a Cohort analysis was performed to further study the success of the assignment in relation to the descriptors considered when subdividing the initial population into groups of similar characteristics. This assessment provided some insights about prone-to-fail submissions---and eventually the students themselves---that could be tackled or, at least, warned in advance to avoid such an issue.

With all above, we may now provide the respective answers to the four Research Questions posed in Section~\ref{sec:method}:

\begin{itemize}
    \item \textbf{RQ\ref{hyp:first}:} \textit{When should students start making submissions to the OJ system?}\\
To answer this particular question we resort to Fig.~\ref{fig:individual_shap} that relates the impact of each submission with respect to the output variable of the model. In this regard, it may be observed that when the submissions are made up to 7 days before the deadline---variable \textit{Days to deadline}---the results are mostly positive.     
    
    \item \textbf{RQ\ref{hyp:second}:} \textit{How many submissions are reasonable for the success of the assignment?}\\
This second point may be also addressed by resorting to Fig.~\ref{fig:individual_shap}. In this case, attending to the number of submitted attempts to the current date---variable \textit{Submissions to date}---it may be observed that the impact is mostly positive when this descriptor is above 40.    

    \item \textbf{RQ\ref{hyp:third}:} \textit{Which would be the risk groups?}\\
To adequately solve this question and identify the risk groups we resort to the Cohort study presented in Section~\ref{subsub:cohort}. In that analysis, one of the particular cohort criteria obtained reported a statistically significant influence on the outcome of the task. This group is the one represented by the condition \textit{\review{`Days to deadline' $<=$ 2.15 \& `Days to deadline'$<$=0.25 \& Submissions to date'$<$=37.5}}, which stands for students who depict a scarce amount of submitted codes in a date close to the deadline of the assignment. As reported in the aforementioned section, elements within this cohort generally lead to not passing the assignment, being hence a clear risk group that should be considered by the instructor of the course.

    \item \textbf{RQ\ref{hyp:fourth}:} \textit{Are there any pieces of advice for the students to adequately address the assignments?}\\
In contrast to the previous questions, to answer this point we require a general vision of the different analyses performed in the work. As commented, students delivering the submissions earlier than a week---\textit{Days to deadline} indicator---who accumulate more than 40 submissions---variable \textit{Submissions to date}---typically lead to positive outcomes, disregarding the particular assignment addressed---\textit{Problem} descriptor in Fig.~\ref{fig:general_shap}. Hence, the student is recommended to start working on the assignment as soon as possible, not to desist even if a relatively large number of submissions are incorrect, and not to discourage due to the difficulty of the task.
\end{itemize}

% -------------------------------------------------------------------------------------------
\section{Conclusions and future work}
\label{sec:conclusions}

Online Judge (OJ) systems have been largely considered in the context of programming-related courses as they provide fast and objective assessments of the code developed and submitted by the students. Despite their clear advantages, OJ systems do not generally provide the student nor the instructor with any feedback from the actual submission besides whether the provided code successfully accomplished the assignment. While this limitation is acceptable up to some extent, it would be useful for these systems to retrieve additional pieces of information that could eventually lead to the identification of student habits, patterns of behaviour, or profiles related to the success (or failure) of the task, among others. Note that, while such types of insights are deemed as key points in the educational field, the process is not currently addressable by existing OJ-based methodologies. 

This work aims to tackle this limitation by resorting to the Educational Data Mining (EDM) field. For that, the proposal considers the use of learning-based schemes from the EDM area---more precisely, Multi-Instance Labelling (MIL) and classical Machine Learning (ML) formulations---to model the student behaviour based on the code submissions provided. In addition, since these frameworks do not generally provide a human-understandable feedback---which is the expected output of the method---, we propose the use of Explainable Artificial Intelligence (XAI) to obtain such interpretable feedback.

This methodology has been evaluated considering a case of study with data gathered from a programming-related course in a Computer Science degree. This collection comprises the different submissions to an OJ system of two different assignments during three academic years, comprising more than 2,500 submissions from roughly 90 different students, which represents all pupils developing the commented course and approximately, 80\% of the people enrolled in it. The results obtained validate the proposal: in terms of statistical significance, the model is capable of adequately predicting the user outcome (either passing or failing the assignment) solely based on the behavioural pattern inferred by the submissions provided. Moreover, the proposal is able to identify prone-to-fail student groups, being hence possible to provide feedback to both the student and the instructor.

Future work considers the further validation of the model, both increasing the amount of data of the case of study as well as considering other alternative courses that also resort to OJ evaluation methods. In addition, we will consider the possibility of exploring the use of human factor characteristics drawn from, for instance, personality, self-efficacy, and motivation tests to boost the prediction accuracy of the system.

\section*{Acknowledgments}
We would like to thank the students who participated in our study. This work has been partially funded by the ``Programa Redes-I3CE de investigacion en docencia universitaria del Instituto de Ciencias de la Educacion (REDES-I3CE-2020-5069)'' of the University of Alicante. The third author is supported by grant APOSTD/2020/256 from ``Programa I+D+i de la Generalitat Valenciana''.

\ifCLASSOPTIONcaptionsoff
  \newpage
\fi



\bibliographystyle{IEEEtran}
\bibliography{referencias}


\begin{IEEEbiography}[{\includegraphics[width=1in,height=1.25in,clip,keepaspectratio]{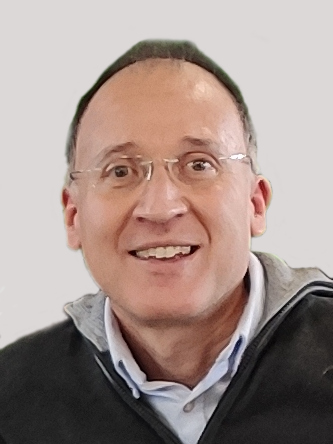}}]{Juan Ram{\'o}n Rico-Juan}
is a Doctor in Computer Science Engineering (2001). Regarding his experience as a researcher, his main interest is related to pattern recognition and machine learning (learning with structured data, editing distances, selection and generation of prototypes, deep neural networks, etc.) in which I have participated in 8 national projects that have resulted in 28 publications in high impact journals (JCR) and 17 in international conferences.
\end{IEEEbiography}

\begin{IEEEbiography}[{\includegraphics[width=1in,height=1.25in,clip,keepaspectratio]{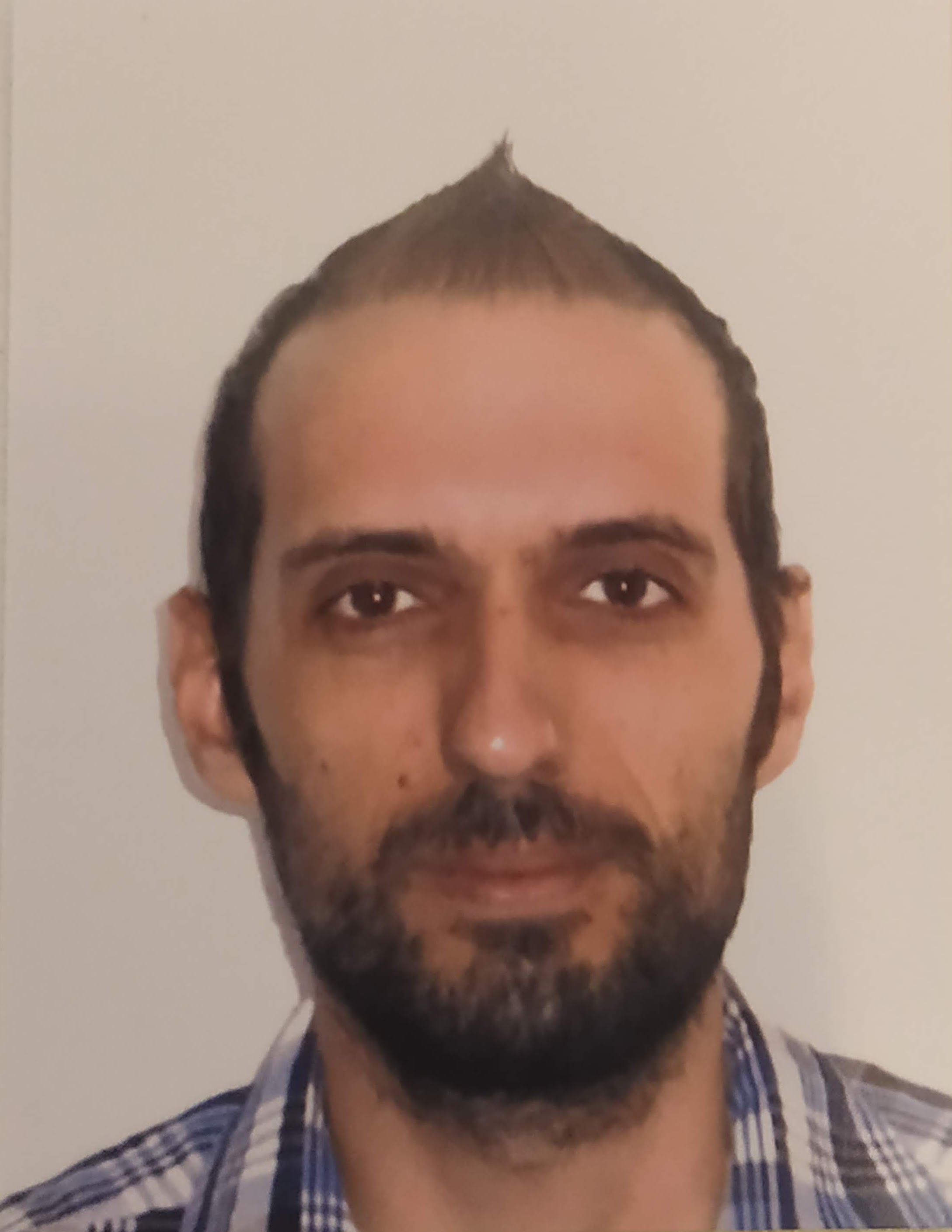}}]{V{\'i}ctor M. S{\'a}nchez-Cartagena}
obtained his PhD in Computer Science in 2015. Since 2020, he has been an Assistant Professor at the Department of Software and Computing Systems at Universitat d’Alacant, Spain. His main fields of research are deep learning and machine translation with an emphasis on low-resource languages and hybrid approaches that combine multiple systems and/or sources of information. He is the author of more than 10 indexed publications, which include top conferences for natural language processing.
\end{IEEEbiography}

% IEEEbiography
\begin{IEEEbiography}[{\includegraphics[width=1in,height=1.25in,clip,keepaspectratio]{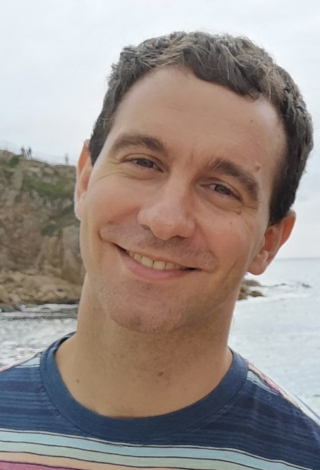}}]{Jose J. Valero-Mas}
obtained the M.Sc. in Telecommunications Engineering from the University Miguel Hernández of Elche in 2012, the M.Sc. in Sound and Music Computing from the Universitat Pompeu Fabra in 2013, and the Ph.D. in Computer Science from the University of Alicante in 2017. He is currently a postdoctoral researcher with a grant from the Valencian Government at the Department of Software and Computing Systems of the University of Alicante, Spain. His research interests include Pattern Recognition, Machine Learning, Music Information Retrieval, and Signal Processing for which he has co-authored more than 30 works within international journals, conference communications, and book chapters.
\end{IEEEbiography}

\begin{IEEEbiography}[{\includegraphics[width=1in,height=1.25in,clip,keepaspectratio]{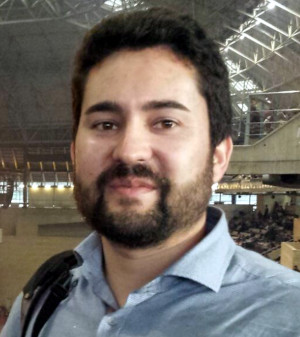}}]{Antonio Javier Gallego}
is an associate professor in the Department of Software and Computing Systems at the University of Alicante, Spain. He received B.Sc. \& M.Sc. degrees in Computer Science from the University of Alicante in 2004, and a PhD in Computer Science and Artificial Intelligence from the same university in 2012. He has been a researcher on 15 research projects funded by both the Spanish Government and private companies. He has authored more than 60 works published in international journals, conferences and books. His research interests include Deep Learning, Pattern Recognition, Computer Vision, and Remote Sensing.
\end{IEEEbiography}


\vfill


\end{document}